\documentclass[a4paper]{jpconf}
\usepackage{graphicx}
\begin{document}
\title{The NEW detector: \\ construction, commissioning and first results}

\author{M Nebot-Guinot for the NEXT Collaboration}

\address{Instituto de F\'isica Corpuscular (IFIC), CSIC \& Universitat de Val\`encia \\Calle Catedr\'atico Jos\'e Beltr\'an, 2, 46980 Paterna, Valencia, Spain}

\ead{miquel.nebot@ific.uv.es}

\begin{abstract}
NEXT (Neutrino Experiment with a Xenon TPC) is a neutrinoless double-beta (\(\beta\beta0\nu\)) decay experiment at the Canfranc Underground Laboratory (LSC). 
It seeks to detect the \(\beta\beta0\nu\) decay of Xe-136 using a high pressure xenon gas TPC with electroluminescent (EL) amplification.\\
The NEXT-White (NEW) detector, with an active xenon mass of about 10 kg at 15 bar, is the first NEXT prototype installed at LSC. It implements the NEXT detector concept tested in smaller prototypes using the same radiopure sensors and materials that will be used in the future NEXT-100, serving as a benchmark for technical solutions as well as for the signal selection and background rejection algorithms. NEW is currently under commissioning at the LSC.\\
In this poster proceedings we describe the technical solutions adopted for NEW construction, the lessons learned from the commissioning phase, and the first results on energy calibration and energy resolution obtained with low-energy radioactive source data.
\end{abstract}

\section{Introduction}
Energy resolution and background suppression are the two key features of any neutrinoless double beta decay experiment. Prototype detectors of the NEXT concept have been used to demonstrate energy resolution better than 1\% FWHM at $Q_{\beta\beta}$ \cite{Alvarez:2012yxw} as well as to demonstrate the background rejection power of track reconstruction in an electroluminescent xenon gas TPC \cite{Ferrario:2015kta}.\\

\subsection{Detection process:}
\begin{itemize}
\item A $^{136}Xe$ nucleus decays emitting the two electrons and both excite the Xe producing primary scintillation light (S1) emission in VUV ($\sim175 nm$) and ionization pairs. 
\item The S1 signal is detected in the PMTs giving $t_0$ and  event position along drift.
\item Electrons create ionization charge in Xe  ($\sim25 eV$ to create one electron-ion pair). 
\item The created electrons drift towards the  anode with velocity $\sim1mm/\mu s$ in a $\sim0.5 kV/cm$ electric filed. 
\item Secondary ionization light (S2) is produced between the gate and anode via the process of EL. The light produced is detected by SiPMs behind the anode used to reconstruct the event topology, or in the PMTs behind the cathode used for the energy measurement.
\end{itemize}

NEW is a step forward in the validation of the roadmap to build the NEXT-100 detector \cite{neus, Martin-Albo:2015rhw}.
It is a 1:2 scale detector using already the facilities, platform, gas system and external shielding to be used for NEXT-100. The construction and commissioning also exercise the assembly procedure, safety and emergency standards, as well as several new technical solutions not tested in the smaller R\&D devices:
cylindrical field cage with a quartz anode plate performing the Time Projection Chamber (TPC),  pressure isolated PMT's, Kapton Dice Boards, readout electronics, criorecovery system and slow controls.

\section{The NEW detector }
Inside the titanium-steel alloy pressure vessel, the TPC is surrounded by radio-pure copper, 6 cm in the barrel and 12 in the endcaps. The  vessel endcaps support the two separate sensor planes, one optimized for calorimetry and the other for tracking. In order to maintain the uniformity of the electric field the High Density Polyethylene  field cage is lined with copper rings connected by a resistor chain. Cathode and the EL region gate are formed by wired grids with the anode being a fused-silica plate coated with ITO for conduction and TPB to convert the VUV light to blue (WLS). The inner part of the field cage hosts a teflon light tube also coated with TPB.The high voltage feedthrough has been designed taking into account the requirement to operate the cathode at up to 50 kV and the grid at up to 20 kV.

\subsection{NEW: Energy plane}
As can be seen in figure \ref{EP}, 12 PMTs (R11410-10 Hamamatsu) coupled with optical gel to sapphire windows coated with TPB to withstand the pressure will operate in vacuum. A thick copper plate holds and separates the PMT-vacuum region from the high pressure region in the chamber, and acts also as a radiation shield. The electronic bases are pinned behind the PMT with heat dissipation connectors and drive the signal outside the chamber.  
\subsection{NEW: Tracking plane}
\begin{figure}[h]
\begin{minipage}{18pc}
\begin{center}
\includegraphics[width=14pc]{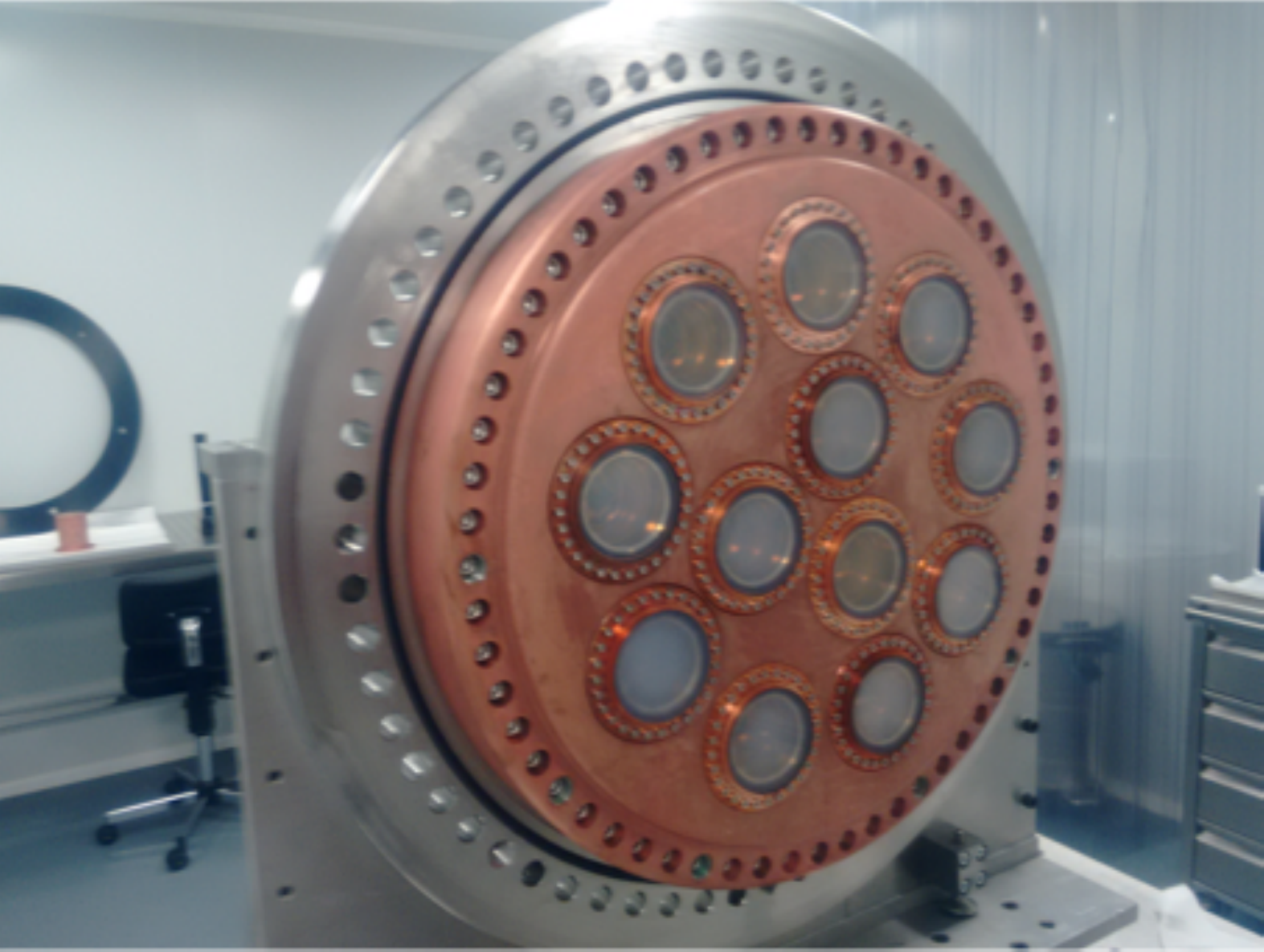}
\caption{\label{EP}Energy plane of NEW.}
\end{center}
\end{minipage}\hspace{2pc}%
\begin{minipage}{18pc}
\begin{center}
\includegraphics[width=14pc]{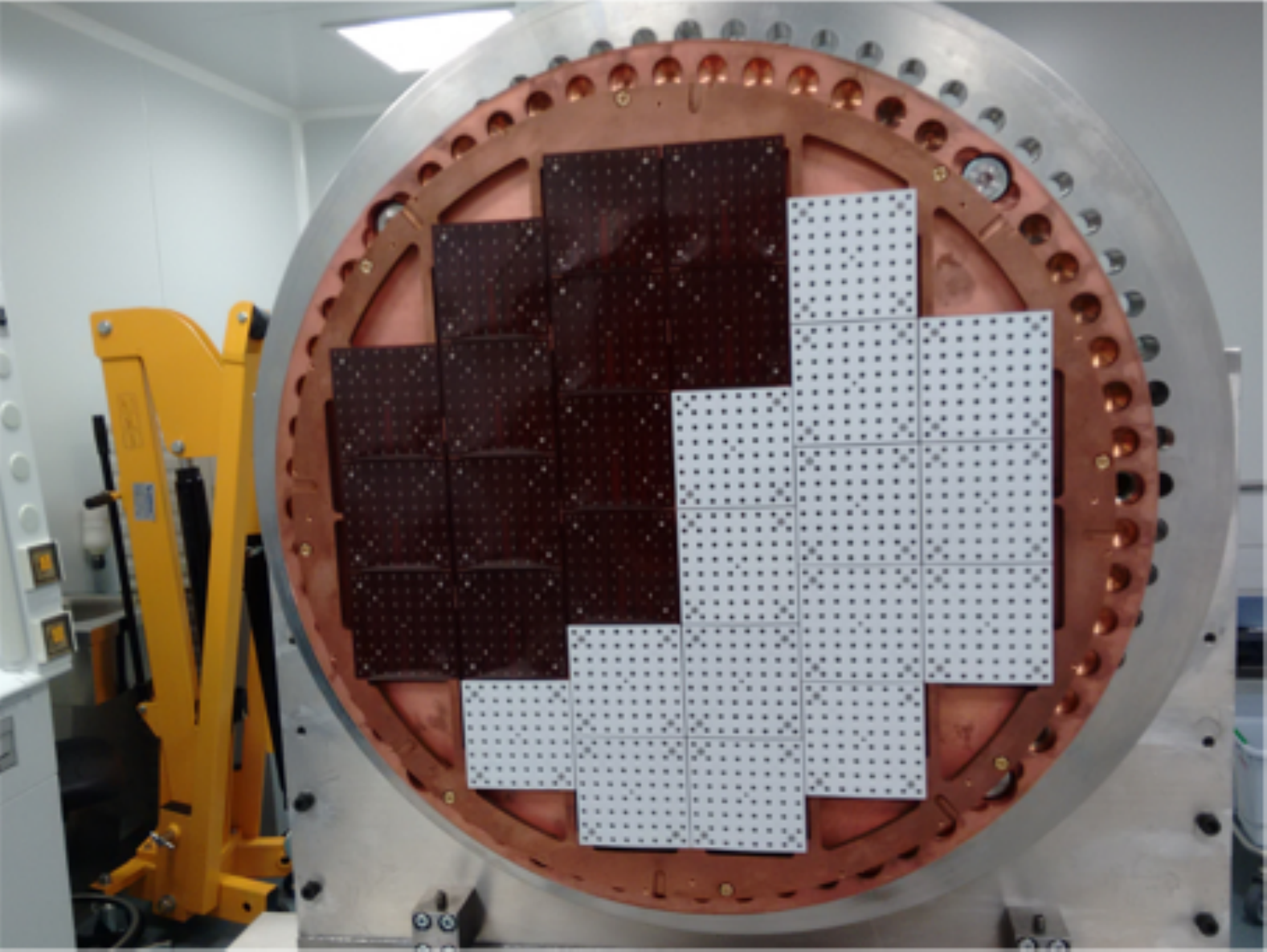}
\caption{\label{TP}Tracking plane of NEW.}
\end{center}
\end{minipage} 
\end{figure}

In figure \ref{TP} is shown the NEW tracking plane made of 28 custom designed Kapton DICE Boards (KDB) with 64  (1x1mm2  SensL) SiPMs each, providing a dense array of 1 cm pitch for topological reconstruction \cite{paola}. 
A reflective teflon mask is fitted on top of each KDB. The KDB built-in tail drives the signal through the copper plate to several gold-ceramic 40 pins feedthroughs at the vessel endcap. All the electronics ( Front-Ends, power supplies, slowcontrols ) stands close to the vessel but outside the shielding castle to reduce the signal transfer mitigation. Also the gas system is embedded to the platform to reduce the dead volumes.  

\section{Calibration} 
\subsection{Sensor calibration}
\begin{figure}[h]
\begin{minipage}{18pc}
\begin{center}
\includegraphics[width=14pc]{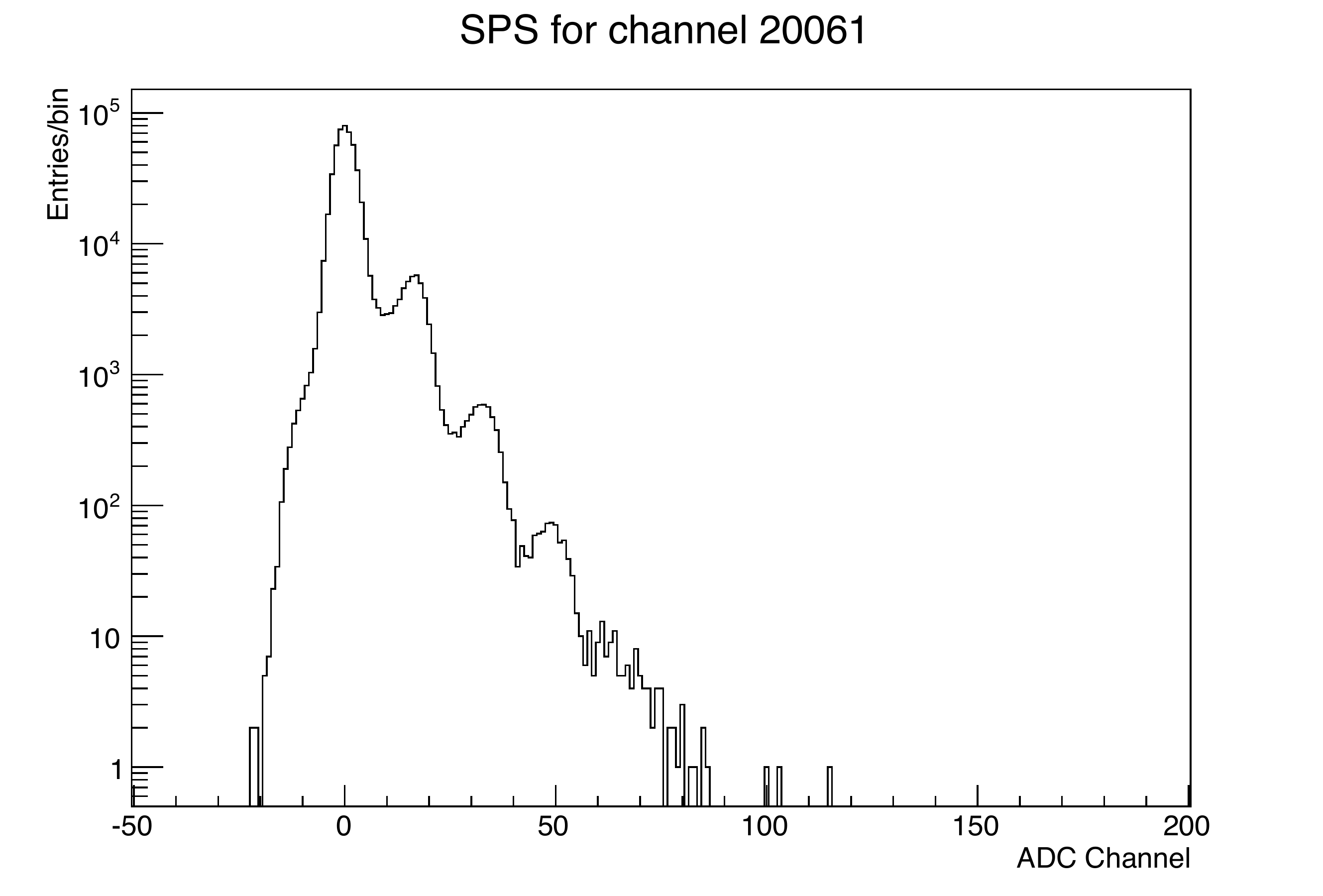}
\caption{\label{SiPMcal} Single Photon Spectrum of a SiPM obtained with dark current events in NEW.}
\end{center}
\end{minipage}\hspace{2pc}%
\begin{minipage}{18pc}
\begin{center}
\includegraphics[width=14pc]{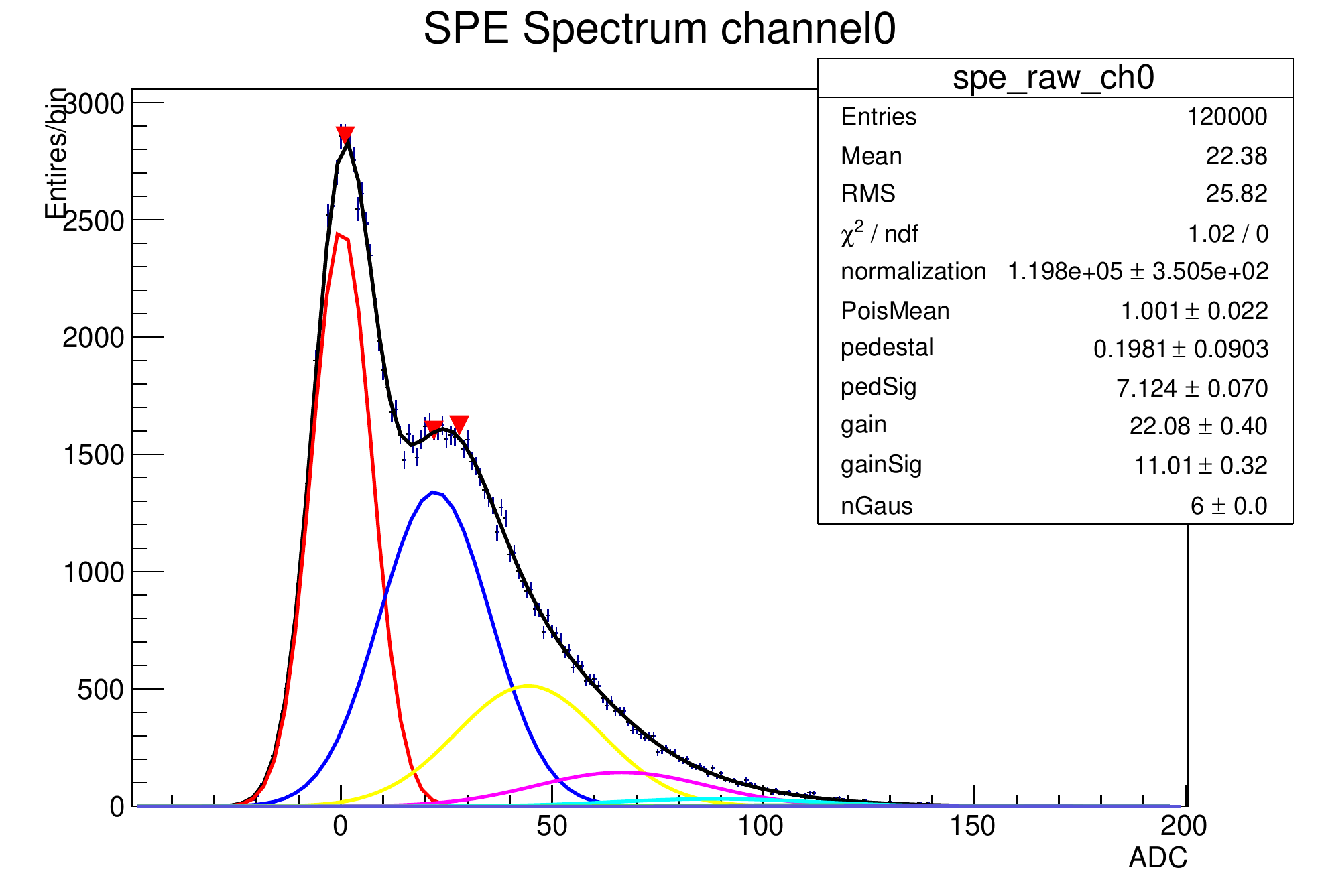}
\caption{\label{SPE}SPE calibration of a central PMT in NEW.}
\end{center}
\end{minipage} 
\end{figure}

The SiPMs are calibrated using the dark current of the sensors. Fitting a Gaussian to the peaks corresponding to the numbers of photoelectrons and using the position in a linear fit provides the conversion gain (fig. \ref{SiPMcal}). The PTMs are calibrated with the single photoelectron (SPE) response by LED light (fig. \ref{SPE}).
\subsection{Energy calibration}
\begin{figure}[h]
\begin{minipage}{18pc}
To demonstrate good energy resolution at higher energies, several calibration source runs will be used: $^{83}Kr$ as a point like source to obtain geometrical correction factors \cite{Lorca:2014sra}, different photopeaks of $^{22}Na$ and $^{208}Tl $ to extract the energy resolution scale. With Monte Carlo studies of $^{208}Tl $ (fig. \ref{ER})  the energy resolution in NEW extrapolates to 0.58\% FWHM at $Q_{\beta\beta}$.
\end{minipage}\hspace{2pc}%
\begin{minipage}{16pc}
\begin{center}
\includegraphics[width=14pc]{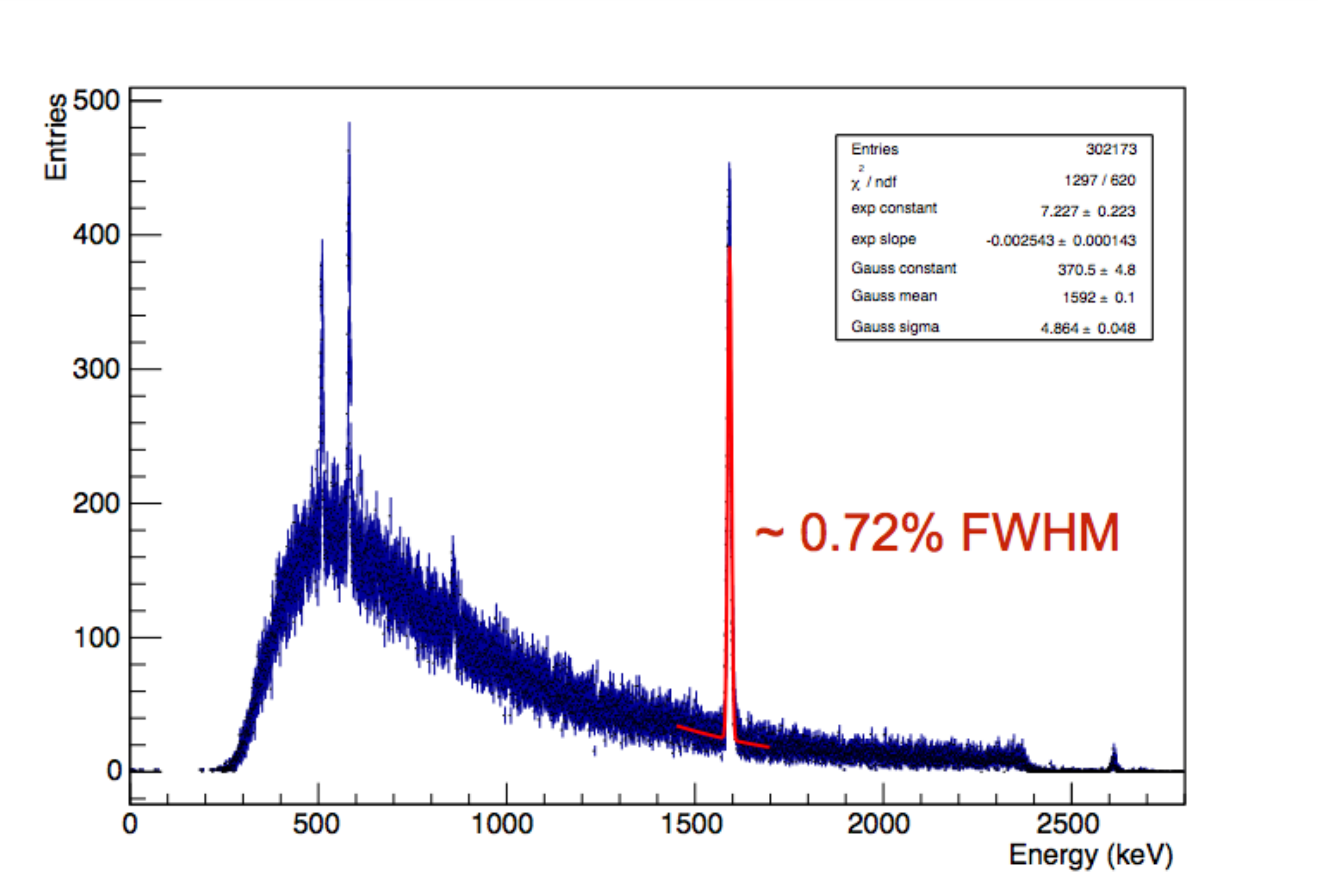}
\caption{\label{ER} MC track energy of $^{208}Tl $ calibration events in NEW.}
\end{center}
\end{minipage} 
\end{figure}

\ack{}
The author acknowledges support from:
the ERC under the Advanced Grant 339787-NEXT; 
the Ministerio de Econom\'ia y Competitividad of Spain and FEDER under grants CONSOLIDER-Ingenio
2010 CSD2008-0037 (CUP), FIS2014-53371-C04 and the Severo Ochoa Program
SEV-2014-0398; GVA under grant PROMETEO/2016/120. Fermilab is operated by Fermi Research Alliance, LLC under Contract No. DE-AC02-07CH11359 with the U.S. DOE



\section*{References}
\bibliographystyle{iopart-num}
\bibliography{biblio}

\end{document}